\documentclass[conference]{IEEEtran}
\IEEEoverridecommandlockouts
\usepackage{cite}
\usepackage{amsmath,amssymb,amsfonts}
\usepackage{algorithmic}
\usepackage{graphicx}
\usepackage{textcomp}
\usepackage{xcolor}
\usepackage[colorinlistoftodos]{todonotes}

\def\BibTeX{{\rm B\kern-.05em{\sc i\kern-.025em b}\kern-.08em
    T\kern-.1667em\lower.7ex\hbox{E}\kern-.125emX}}

\newcommand\copyrighttext{%
  \footnotesize \textcopyright 2023 IEEE. Personal use of this material is permitted. Permission from IEEE must be obtained for all other uses, in any current or future media, including reprinting/republishing this material for advertising or promotional purposes, creating new collective works, for resale or redistribution to servers or lists, or reuse of any copyrighted component of this work in other works. 
  \linebreak Preprint submitted to the 21st IEEE International Conference on Industrial Informatics, INDIN'23.
  }
\newcommand\copyrightnotice{%
\begin{tikzpicture}[remember picture,overlay]
\node[anchor=south,yshift=10pt] at (current page.south) {\fbox{\parbox{\dimexpr\textwidth-\fboxsep-\fboxrule\relax}{\copyrighttext}}};
\end{tikzpicture}%
}

\begin{document}

\title{A Design Approach and Prototype Implementation for Factory Monitoring Based on Virtual and Augmented Reality at the Edge of Industry 4.0
\thanks{This work has been partially supported by the EU research project ENERMAN (ENERgy-efficient manufacturing system MANagement), funded by the European Commission under H2020 and contract number 958478.}
}
\makeatletter
\newcommand{\linebreakand}{%
  \end{@IEEEauthorhalign}
  \hfill\mbox{}\par
  \mbox{}\hfill\begin{@IEEEauthorhalign}
}
\makeatother

\author{\IEEEauthorblockN{Christos Anagnostopoulos\thanks{$^*$C. Anagnostopoulos and G. Mylonas have contributed equally.}}
\IEEEauthorblockA{\textit{Industrial Systems Institute} \\
\textit{Athena Research \& Innovation Center}\\
Patras, Greece \\
ORCID: 0000-0002-7998-5708}
\and
\IEEEauthorblockN{Georgios Mylonas}
\IEEEauthorblockA{\textit{Industrial Systems Institute} \\
\textit{Athena Research \& Innovation Center}\\
Patras, Greece \\
ORCID: 0000-0003-2128-720X}
\and
\IEEEauthorblockN{Apostolos P. Fournaris}
\IEEEauthorblockA{\textit{Industrial Systems Institute} \\
\textit{Athena Research \& Innovation Center}\\
Patras, Greece \\
ORCID: 0000-0002-4758-2349}
\linebreakand
\IEEEauthorblockN{Christos Koulamas}
\IEEEauthorblockA{\textit{Industrial Systems Institute} \\
\textit{Athena Research \& Innovation Center}\\
Patras, Greece \\
ORCID: 0000-0001-7172-0628}
}

\maketitle

\copyrightnotice

\begin{abstract}
    Virtual and augmented reality are currently enjoying a great deal of attention from the research community and the industry towards their adoption within industrial spaces and processes. However, the current design and implementation landscape is still very fluid, while the community as a whole has not yet consolidated into concrete design directions, other than basic patterns. Other open issues include the choice over a cloud or edge-based architecture when designing such systems. Within this work, we present our approach for a monitoring intervention inside a factory space utilizing both VR and AR, based primarily on edge computing, while also utilizing the cloud. We discuss its main design directions, as well as a basic ontology to aid in simple description of factory assets. In order to highlight the design aspects of our approach, we present a prototype implementation, based on a use case scenario in a factory site, within the context of the ENERMAN H2020 project.
\end{abstract}


\section{Introduction}

Boosting situational awareness inside complex, dynamic environments can allow a better grasp of the current situation, improve decision-making and performance, among other aspects. As such, in recent years we have witnessed a wealth of activity focusing on enabling such dimensions in a number of sectors, and specifically in industrial settings. Having this in mind, the rapid advancement in technologies related to virtual and augmented reality (VR and AR respectively), together with the commercial availability of devices that can enable their use, has facilitated the introduction of said technologies in industrial use-cases and settings. However, since the field is fairly new and the related enabling technologies are rapidly evolving, there is still a lot of room available for new design and implementation approaches, while there is a limited number of established design ``patterns''. There are also several open issues on aspects like e.g., ontologies for describing assets inside a factory space.

At the same time, numerous commercial and research activities are currently aiming to utilize such technologies. Among them, the ENERMAN\footnote{ENERMAN Horizon 2020 project, https://enerman-h2020.eu/} EU research project focuses on energy efficiency on the industrial space, utilising several tools and promote this goal within several different industrial applications. Among its research directions, ENERMAN includes use of VR/AR-based approaches for supporting end-users towards boosting situational awareness and improving energy-related decision-making on the factory floor.

Within this context, we designed and developed a prototype of a ``toolbox'' that offers several blocks for facilitating the creation of real-time visualization inside an industrial space, based on a design targeting both VR and AR, and aiming to improve human-in-the-loop awareness, when being either on-site, or in remote locations inspecting the situation inside a factory. We contribute to the currently ongoing discussion by presenting our approach, reporting on the aims of this toolbox, its design and the user requirements upon which it was based on, its implementation and the technologies utilized to produce this result, together with a brief presentation of a prototype implementation that serves as a showcase for its capabilities, which was loosely based on one of the ENERMAN partners' factories. 

\section{Previous Related Work}

A large volume of work on defining and testing various VR and AR approaches within the context of various applications has surfaced in the past few years. A recent survey \cite{ar-vr-in-industrial-systems-review} regarding AR and VR applications in industrial systems, states this field is rising in popularity, with maintenance and virtual training being the most popular fields of application at that point in time. Among the benefits of using AR/VR techniques in industrial spaces, the authors mention on-site diagnostics, safety control, better training, product reconfiguration, improved Human-Machine Interface (HMI) experience. 

As regards the use of recent HMI technologies within the Industry 4.0 context, \cite{buttner2023} discusses a range of issues and aspects, concluding that the integration of such technologies is still largely an open issue, although we are witnessing great leaps in fields like AR and VR. In addition, questions like how will we create more human-centered workplaces using such technologies and what is the user acceptance of such solutions also require further research.

AR has enjoyed a great deal of attention in the past few years, with work such as \cite{ar-apps-in-industry-4} and \cite{survey-industrial-ar} surveying the use of AR for realizing Industry 4.0 applications. The authors of \cite{ar-apps-in-industry-4} also claim that maintenance is the most popular AR application in an Industry 4.0 context, at least in terms of research work, followed by assembly and human-robot collaboration, and then at a significantly lesser scale by manufacturing and training. They also identify smart glasses and tablets as the devices mostly used in such AR applications. 

In terms of types of industries utilizing AR, the automotive, mechanical and electronics industries top the list included in \cite{survey-industrial-ar}. Regarding devices used for AR applications, the authors report that Head-Mounted Displays (HMD) were the most cited type of device for such AR applications, followed on a lesser scale by tablets and smartphones. In terms of toolkits for AR development, the authors reported that ARToolkit and Vuforia were the two most popular tools in 2020. 

The authors of \cite{user-acceptance-in-industrial-ar} examine the user acceptance of AR in industrial contexts via means of a survey of recent work in the field. Once more, maintenance is reported as being the most popular application area, followed by assembly. They also report that use of HMDs and mobile devices appears to be equally divided in applications in industry, while also mentioned the use of projection-based methods in such applications. Moreover, they point out the deficiencies in, or lack of, consideration of user acceptance aspects in the evaluation of AR-based solutions in Industrial AR so far. 

As regards the use of VR in industrial spaces, \cite{use-of-vr-in-product-design} and \cite{apps-of-vr-in-maintenance} provide insights into the recent related work in the field. The authors of \cite{use-of-vr-in-product-design} reported on the uses of VR available at that time, claiming that utilization in the automotive sector for product design purposes was the most common example of VR’s use in industry, e.g., for evaluating visibility or ergonomics inside a future product, or for enabling storytelling and facilitating communication across teams with different backgrounds (designers, engineers, etc.). A more recent survey \cite{apps-of-vr-in-maintenance}, reports on the application of VR in maintenance during product lifecycle. The authors underline VR’s role in the design process, as well as in maintenance, claiming that VR can play a role in the entire lifecycle of industrial products.

The efficient utilization of manufacturing resources is a major concern due to the significant heterogeneity of the data generated in manufacturing process. 
 Over the last years, the increasing digitization of businesses and the adoption of novel technologies has resulted in technological ecosystems which are comprised of multiple components that can store, process and generate a huge amount data continuously from the shop floor \cite{app11115110}. These data can only be transformed into valuable information and utilized in multiple applications, (e.g. predictive maintenance, monitoring etc.) only if they are gathered and organized in a standardized way. To this end, in \cite{LONGO2022107824}, \cite{LONGO2022594} an ontology-based Industry 4.0-ready architecture is presented, which is compliant to FAIR \cite{wilkinson2016fair} data principles and introduces a novel ontology that aims the semantic and syntactic integration. Finally, in \cite{Inoovas} the authors present Inoovas, which is an ontology that focuses on the authoring of AR and VR operations in the context of a manufacturing environment.

From the above, it is apparent that both AR and VR are at the epicentre of  activity towards implementing the Industry 4.0 vision, but at the same time there is a lot of research on both how and where they should be used within industrial spaces. In this sense, both the range of technologies used in the actual devices available today, as well as the application field in which such devices are employed is still to a large degree a work-in-progress area, although very promising. 

\section{Aims, Architecture \& Components}

As mentioned in the introduction, in this work we report on an AR/VR ``toolbox'', which is meant to facilitate monitoring and decision-making at the factory floor with respect to the goals of the ENERMAN project. In general, the project has several different pilot sites, ranging from the automotive industry to chocolate sweets or aluminum production, in which the project tools will help with detecting processes and situations where energy savings can be implemented. In this context, the AR/VR toolbox is meant to help end-users for on-site surveying and to improve their situational awareness with respect to processes and data related to energy in manufacturing.

Regarding the design of the toolbox, a multitude of aspects had to be considered. First of all, the overall approach aims to be flexible enough to accommodate implementations in diverse factory settings. Moreover, apart from the dimension of using extended reality tools inside the real factory space, during the Covid-19 pandemic we witnessed a growing request for tools that  allow remote inspection of industrial spaces. After conducting an initial survey in the current state-of-the-art in the field, a first set of design principles was used to produce an initial version of the design for the toolbox:

\begin{itemize}
    \item The main design directions should be usable under both VR and AR application settings.
    \item Users should not be overwhelmed with information, but instead visualizations should follow a location and situational awareness-based approach.
    \item Visualizations should utilize empty space where available in the real world, in order to provide a richer layer of information inside industrial spaces.
    \item The visualization approach should follow an approach based on simplicity and economy, while also being aware of the design approach in other ENERMAN tools in order to have a relatively coherent visual identity.
\end{itemize}

Apart from such design principles,  we initially produced three general design ``directives'' as regards what the visual part should provide to the end-user through VR and AR:

\noindent\textbf{Overview of an area's status:} Users can preview the status of an entire area inside the factory by utilizing quick updates, as well as information concerning the whole space. For example, power consumption of the entire area viewed by the user, or the respective environmental data. this feature can be useful to quickly survey the status of an area in the factory.

\noindent\textbf{Information tied to specific machinery/part:} Users can quickly discern the status of a specific part of the production chain when approaching it. When the user is in certain range of the area of interest, information regarding this part appears. E.g., when approaching a cooling tunnel, information regarding the actual power consumption, as well as power consumption predictions made by the ENERMAN software. Another example is when approaching a liquid container, to quickly check on the temperature and contents of the container.

\noindent\textbf{Use proximity for situational awareness:} Users can get information regarding more complex parts of the factory depending on proximity to the area of interest, as well as their actual current view, i.e., get data for parts that are near them or in their field-of-view. E.g., users get information on specific compartments of a cooling tunnel and the actual temperature/humidity data as they change their view. The system also provides notifications about alerts generated e.g., when temperature inside a specific compartment goes above  a certain threshold that is crucial for some production process.

With respect to general functional requirements due to the aims of ENERMAN, the toolbox should:

\begin{itemize}
    \item Visualize data from both the edge and the cloud.
    \item Provide a basis for Digital Twin-related visualization.
    \item Provide generic ``templates'' for developing site-specific implementations in VR/AR.
    \item Be able to run on HMDs, smartphones and tablets.
\end{itemize}

As regards the overall architecture and the ENERMAN software components used by the toolbox as data inputs, it interfaces with two components: a) the Data Aggregator, and b) the Big Data Analytics Engine. A high-level architecture diagram for the toolbox can be seen in Fig.~\ref{fig:architecture}.

\noindent\textbf{Data aggregator (DA):} DA provides the toolbox with close to real-time data from the edge, as well as notifications. In this way, there is a direct link to data at the edge, which in some cases provides an advantage in terms of simplicity and speed/responsiveness, or in other cases due to privacy, security or business reasons, such data cannot leave the factory. Furthermore, the elements of ENERMAN developed at the edge provide data and functionality which is not available on higher levels, related to data anomaly detection, and other data-related operations. Thus, a direct link is required to have a more complete view of the data available at the edge and provide richer functionality to the potential end-users, which in this cases are the pilot sites of the project. The edge can also provide direct access to notifications produced by performing data anomaly detection at the edge, since this kind of functionality is not available for all pilot sites in the project, and especially in a near real-time manner. 

\noindent\textbf{Big Data Analytics Engine (BDAE):} The BDAE provides the toolbox with historical data from the pilots of the project, as well as constantly performs analytics on such data and makes it available to other software components, such as e.g., 2D visualization tools. In this case, the BDAE can provide applications with large volumes of data from the sites of the project. Its API has been developed to facilitate the ENERMAN components accessing historical data per use-case and process, request predictions and analytics on selected data, and download pre-processed data to use for downstream tasks.

\begin{figure}
\begin{center}
    \includegraphics[width=\columnwidth]{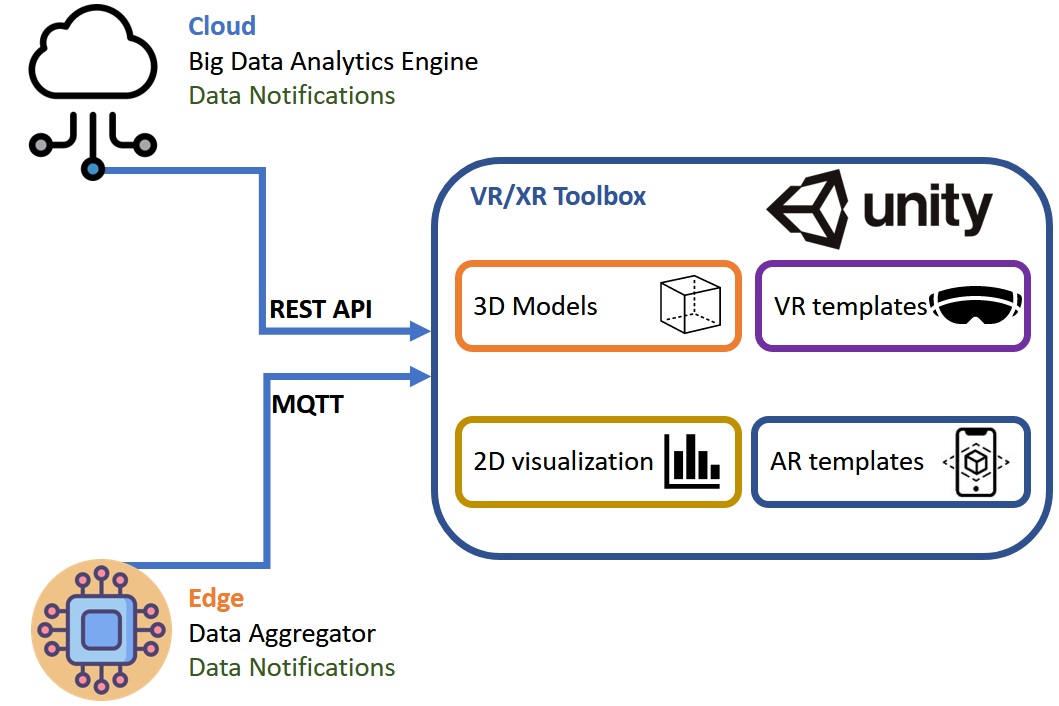}
    \caption{Overview of the system architecture.}
    \label{fig:architecture}
\end{center}
\end{figure}

\section{Implementation Details}

In terms of interfaces, the toolbox utilizes the exposed MQTT and REST APIs respectively, which are presented in more detail later in this section. The toolbox utilizes the Unity platform, which as discussed in the next section, provides a common framework for VR and AR applications, which is also target-agnostic, i.e., it provides us with great flexibility to support a range of different devices without requiring additional development effort. Within the toolbox, a set of ``templates'' have been developed for VR and AR in order to enable the realization of certain experiences in AR and VR for the project. Such templates encompass the design elements presented in previous sections, and also deal with the intricacies of interfacing with other components. 

These ``templates'' exist over a set of 3D models produced by the consortium and 2D visualization capabilities provided by Unity and Unity plugins. The 3D models, used mainly for the VR part of the equation, correspond to entities existing in the pilot sites of the project, i.e., they could be a 3D model of a pilot site, including machinery, HVAC, etc., that can be categorized as belonging to a set of ``resources'' that correspond to such real-world entities.

In order to organize both visually and semantically the information displayed in AR/VR modes, and also describe this information in a machine-friendly manner, we have adopted an ontology (see Fig. ~\ref{fig:ar-description-ontology}) in terms of resource description, based on the following types:

\begin{itemize}
    \item \textit{Site} is the uppermost element in the hierarchy, referring essentially to one pilot site in the project. All of the other resource types live under sites, and it serves to identify the overall factory site in which a certain resource can belong to.
    \item \textit{Departments} refer to different parts of a factory site, mostly in terms of production line but also in terms of physical space (although this varies from site to site, since different production lines can overlap in physical space). Sites can have multiple departments.
    \item \textit{Assets} identify specific items inside a department, such as machines that are important parts of the line inside a department. Assets can also represent entities such as environmental sensors that describe overall values and do not belong to a specific machine. Departments can have multiple assets.
    \item \textit{Resources} identify specific aspects that are important to the overall operation of a specific asset and describe its status. Resources can refer to simple, or more complex aspects, that comprise multiple dimensions for their description.
    \item \textit{Data} are used to describe the actual values corresponding to the description of the Resources’ status. E.g., if a certain machine has an energy consumption meter, it can have multiple data referring to different aspects such as momentary power, predicted consumption, average consumption, etc.
\end{itemize}

\begin{figure*}
\begin{center}
    \includegraphics[width=0.6\linewidth]{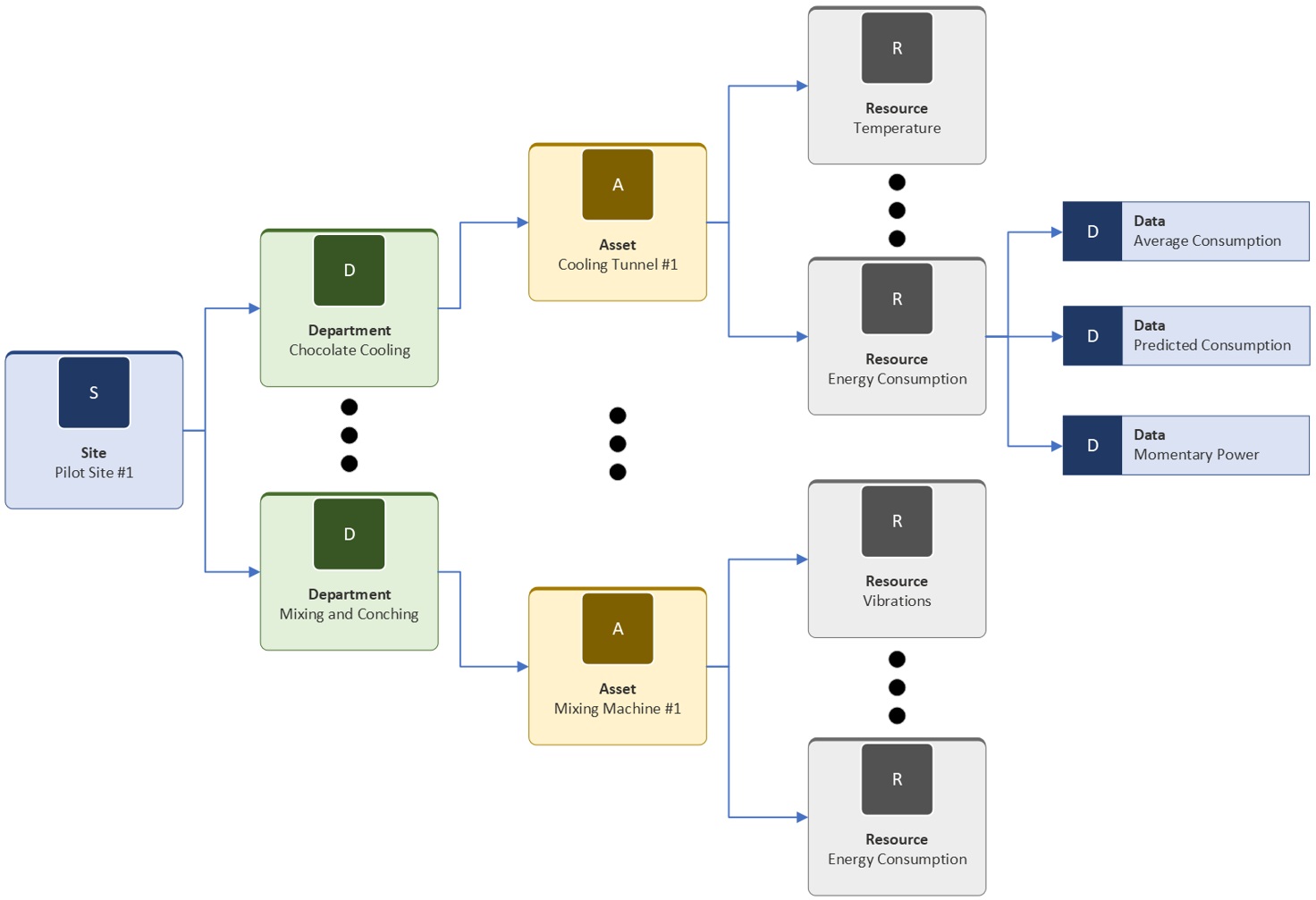}
    \caption{High-level view of an example built using the ontology described in this work.}
    \label{fig:ar-description-ontology}
\end{center}
\end{figure*}

\begin{figure}
\begin{center}
    \includegraphics[width=0.95\columnwidth]{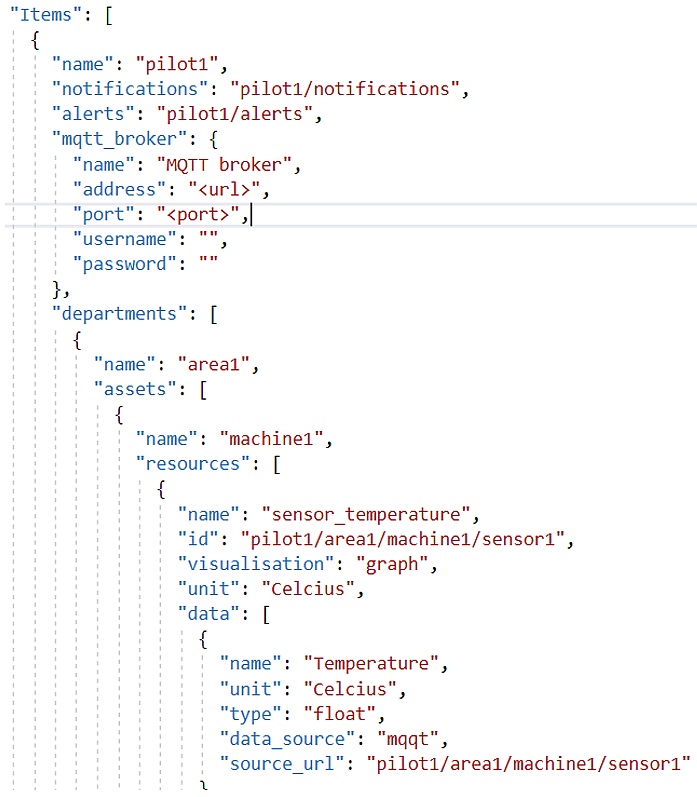}
    \caption{Example of a basic JSON configuration file for the system.}
    \label{fig:json-configuration-example}
\end{center}
\end{figure}

Each of the aforementioned types can have multiple properties in order to describe it, e.g., a name, an identifier, coordinates, and so on. This typology can be used to identify a specific aspect inside an EnerMan factory site, which can then be utilized to both retrieve the actual corresponding information, as well as structure the way information is displayed in a simple, human and machine-readable format. E.g., this makes it straightforward to use a script on top of a representation that uses a JSON-based format (Fig.\ref{fig:json-configuration-example}) to define what is displayed for a specific part of the factory in AR and VR, instead of a monolithic, inflexible implementation.

In terms of 2D visualizations, the toolbox can provide a range of capabilities to the developers. Users are able to see charts and graphs of e.g., energy consumption trends inside the factory, predictions for energy consumption, or graphs for environmental conditions over time, and so on. These are complemented by resource-specific information e.g., for a specific machine in the production line, presented in a tabular format, aiming to provide a quick overview of the status of that specific resource.

\subsection{Interconnection with other EnerMan components}

We now describe in detail the ways in which the toolbox currently utilizes other existing components of the EnerMan software stack, in order to have access to data from the pilot sites. As mentioned above, there are 2 main avenues for this, with the first being through edge node data collection and pre-processing facilities, and the second via the BDAE.

The Edge data collection and pre-processing embedded platform implemented supports multiple upstream protocols, such as MQTT and OPC UA, in order to transfer stream or batch data to the other subsystems or components of the ENERMAN architecture. In the case of interfacing the VR/AR toolbox with this component, we have chosen to use MQTT, due to its flexibility for setting up connections and ease of interfacing with client applications, such as the case described here.

For the sake of demonstrating the capabilities of this component and testing our implementation, we set up a ``demo'' pilot site, based on the actual setup of a part of the production line of an ENERMAN pilot site. In this context, this pilot site provides data for the following aspects:

\begin{itemize}
    \item Environmental conditions of the overall physical space of the site, in the form of average temperature and relative humidity conditions.
    \item Readings about the real time status of 4 liquid containers, as regards the temperature and fullness of each tank.
    \item A set of cooling tunnels provides energy consumption readings for each tunnel.
    \item Based on actual energy consumption data from a specific ENERMAN pilot site, we created 3 datasets based on different energy prediction algorithms developed. Thus, the actual energy consumption readings are paired with energy predictions at the moment of each actual energy reading.
    \item To simulate the case that a data anomaly detection algorithm is executed at the edge, and it encounters certain anomalies during operation, we created a set of events that trigger the creation of certain notifications at the VR/AR client side. 
\end{itemize}

For this ``demo'' use case, and having the above in mind, the edge nodes have been configured to provide structured data in JSON format through the MQTT protocol and a set of sample topics that were used to validate the functionality of the toolbox and demonstrate its capabilities.

\subsection{VR/AR components and Prototype Implementation}

Unity\footnote{Unity, Real-time development platform, https://unity.com/} was used for the development of the VR/AR components. It is an industry-leading software development platform with wide community support. Our purpose is to develop technology-agnostic components that can be ported to different platforms. Unity's XR Plug-in Framework is a unified plug-in framework enabling direct integration to different platforms. The notion behind this framework is a common API provided by Unity that exposes common functionalities to the supported platforms. Thus, each software or hardware provider can develop their own Unity plug-ins. 

As regards the technological side, apart from the Unity XR framework, one of the biggest advantages of the Unity ecosystem is that it offers a wealth of readily available plugins that aid developers in their implementations, that may require specialized views for which considerable implementation effort could otherwise be required. Specifically for displaying charts inside Unity, there is currently a number of commercial plugins offered in Unity’s Asset Store, the online platform for Unity developers’ resources. For our prototype implementation, we have picked the ``Graph and Chart''\footnote{Graph and Chart, Bitsplash IO, http://bitsplash.io/graph-and-chart} Unity plugin, which offers a range of visualization options for graphs and charts, and it is also compatible with both VR and AR development in Unity. As such, it greatly simplifies the development of visualizations for depicting the data collected from the EnerMan infrastructure, and in this case the prototype implementation for the toolbox. Furthermore, since the visualization and application logic are completely separated in the implementation of the toolbox, a different plugin could be utilized should the end-user or pilot site choose to do so, or in the case where different kinds of visualization approaches are required.

\subsection{VR components}

In the following figures, some of the assets created for the visual and functional representation of the real-world assets inside Unity are illustrated. The development of the models follows a pipeline which contains the following steps:

\begin{itemize}
    \item \textit{Data gathering}: in the case of LiDAR scanning or photogrammetry, the accurate and precise create of a dataset can lead to the successful generation of 3D models.
    \item \textit{3D modelling}: the product of the previous step, or a completely new object,  edited in either a third-party modelling tool or in Unity.
    \item \textit{Component Development}: takes place inside Unity and comprises the selection and development of properties and behaviours that capture more accurately the functional characteristics of the real-world object.
\end{itemize}

As regards the translation of the design elements presented earlier, we continue with the presentation of some characteristic examples, as they were implemented inside Unity. Such 3D objects can be manipulated programmatically inside Unity, giving developers easy access to modifying their characteristics and adding custom visualizations over them in real time, responding to data from other components and interaction with the end-user.

With respect to presenting an overview of an area's status, Fig.~\ref{fig:vr-example-1} shows a screenshot where the user views an industrial space and overall power consumption trends are projected on top of a wall, thus utilizing the empty area available and continuously showing data without either blocking the user's view of other aspects, or requiring other user interaction. Fig.~\ref{fig:vr-example-2} gives another example of a visualization in front of a cooling tunnel, thus providing information tied to specific machinery/part. When the user approaches the machine in virtual space within a given distance, a visualization displaying data about power consumption is presented to the user, hovering over the cooling tunnel. This is also a good example of the same design approach that can be used in both AR and VR scenarios, since we are using the same technical basis as mentioned earlier in this section. Fig.~\ref{fig:vr-example-3} presents another example of using proximity for situational awareness, where the user can see information regarding specific compartments of the tunnel, i.e., with an additional degree of location awareness, while also being able to see the overall power consumption figures in the background.

As a side note, the current implementation of the prototype in Unity is ``light'' enough so as to be able to run inside an Oculus Quest 2 HMD, which essentially has the processing power of a current mid-range smartphone.

\begin{figure}
\begin{center}
    \includegraphics[width=\columnwidth]{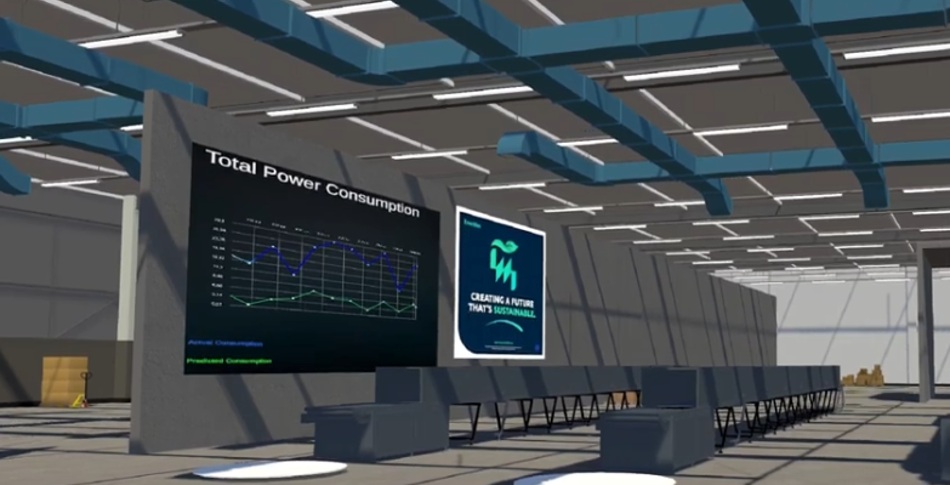}
    \caption{Visualization in VR, an example of an overview of data for an area's status.}
    \label{fig:vr-example-1}
\end{center}
\end{figure}

\begin{figure}
\begin{center}
    \includegraphics[width=\columnwidth]{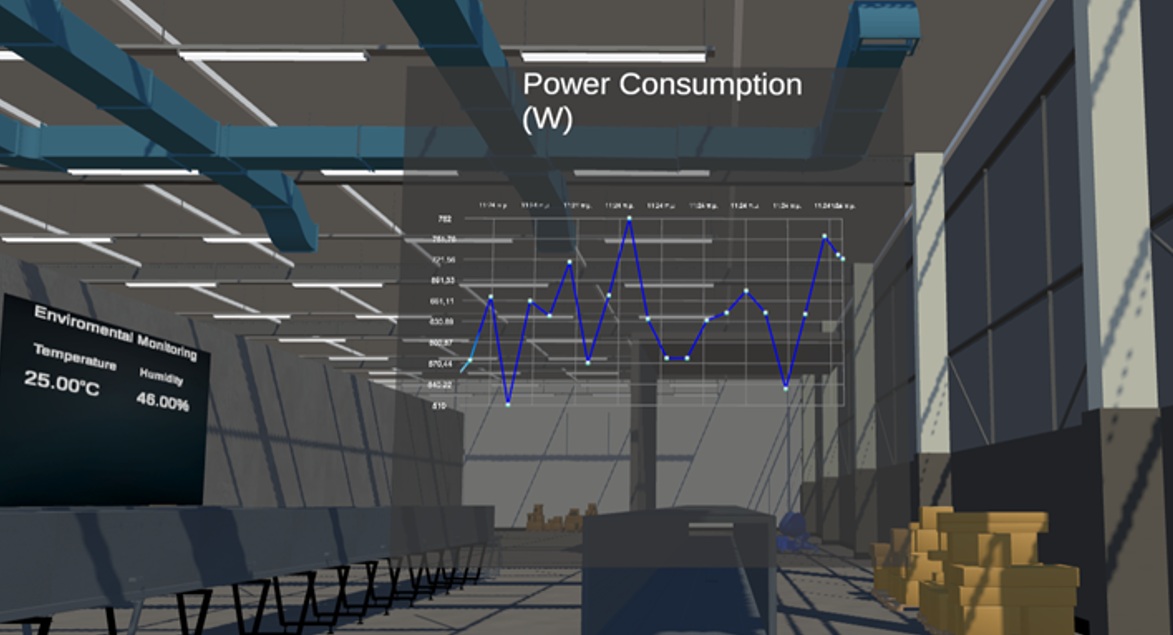}
    \caption{Visualization in VR, an example of information tied to specific machinery/part and using proximity for situational awareness.}
    \label{fig:vr-example-2}
\end{center}
\end{figure}

\begin{figure}
\begin{center}
    \includegraphics[width=\columnwidth]{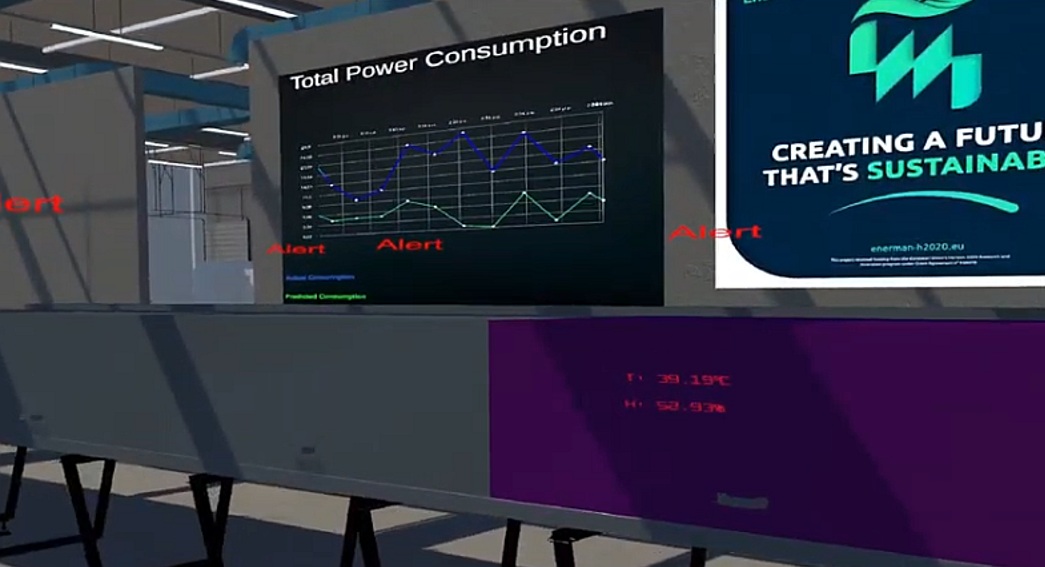}
    \caption{Visualization in VR, an example of using proximity for situational awareness.}
    \label{fig:vr-example-3}
\end{center}
\end{figure}

\subsection{AR components}

We continue with the description of some of the central AR aspects. As mentioned previously, we utilize the ``Graph and Chart'' plugin to generate visualizations for the data received from the Edge and the Cloud, using the API offered by the respective components. Beginning with a basic AR example, in the following figure you can see an example of a chart visualization in AR mode, using this plugin and displaying data received directly from intelligent edge components.
 
Moving on to some more practical examples of the use of AR in the project, in the following figures you can see examples of the utilization of AR in a basic example use case that can be encountered: a power distribution panel that has no real-time information visualization capabilities, but for which it could be useful to have an AR view displaying related data about current energy consumption or average statistics. For this specific example, we showcase in Fig.~\ref{fig:ar-panel-example} a hierarchical display of related information:

\begin{itemize}
    \item The power distribution box has a cover, which can be opened to reveal the innards of the box.
    \item Information is presented in a tabular form in both hierarchical layers for readability purposes.
    \item In the left part of the following figure, we showcase what EnerMan end-users can see on the exterior of the box, i.e., data about the current status of the power distribution box and related notifications. The status consists of a general characterization about the system (``normal'') and a reading of the current power distributed (``5.7kW''). Underneath the status data, most recent notifications can be displayed (2 of them displayed in this case).
    \item In the right part of the figure, the box is opened, and the user can now see more information regarding the actual layout of the power distribution box panel, where energy meters monitor discrete parts of the box. The AR data displayed reveal information about each meter, its status (``OK'' or ``ATT'', for attention), and momentary power readings (``1.3kW'', etc.).
\end{itemize}

\begin{figure}
\begin{center}
    \includegraphics[width=\columnwidth]{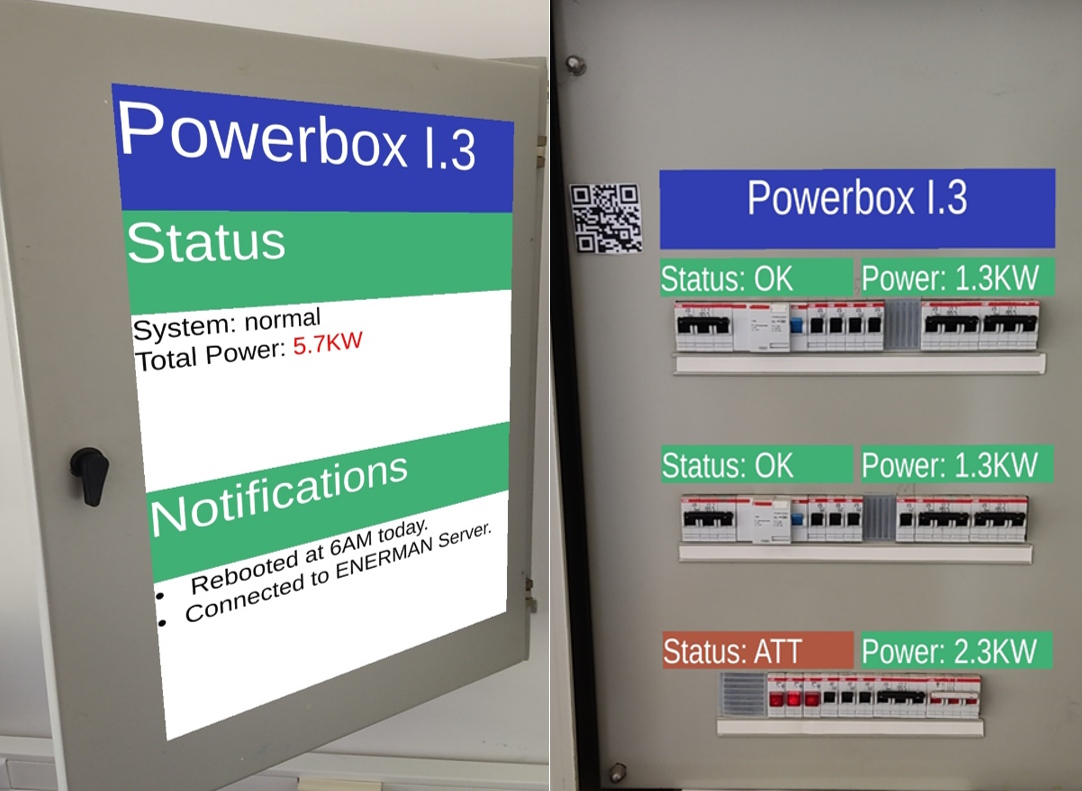}
    \caption{Example of a 2-level AR-based visualization of a POI at an electricity distribution panel: on the exterior, more general information is displayed, while on the interior more specific info about the status of the system.}
    \label{fig:ar-panel-example}
\end{center}
\end{figure}

\begin{figure}
\begin{center}
    \includegraphics[width=\columnwidth]{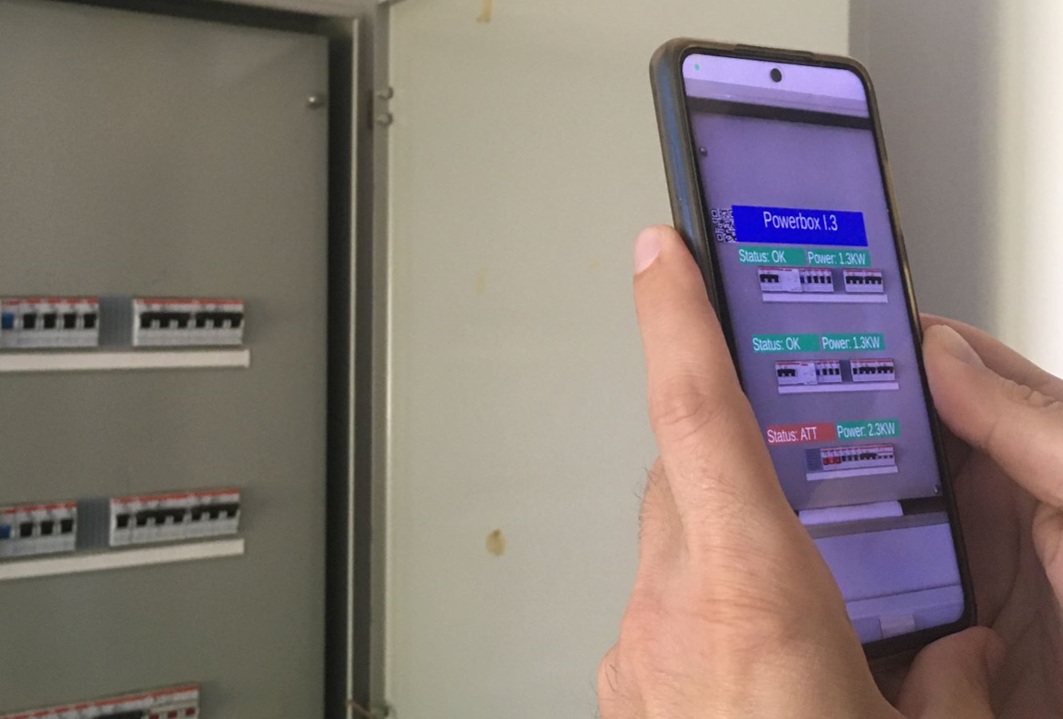}
    \caption{View of the AR interface in action when using a smartphone.}
    \label{fig:holding-phone-view}
\end{center}
\end{figure}

\subsection{Demonstration prototype}

We implemented a prototype ``pilot'' site to drive the design and development of the toolbox, especially as regards VR aspects, but also to help develop the design elements that were to a certain applied in the AR developments as well. Since we have presented the design and implementation elements in previous section, we will briefly present here the process followed to develop the prototype, in order to give insights as to how future developments for the implementation of virtual factory components and digital twins for specific pilot sites in the project will unfold. In order to have a foundation for the design and development that bears to a certain degree resemblance to the layout and organization of a factory, we decided at an early stage that such a prototype should be to an extent similar to one of the pilots in the project. 

For kick-starting this process, we visited one of the pilot sites in order to see and understand the basic parts involved in the production lines that are related to the ENERMAN project and the respective scenarios. After discussion with the respective consortium partner, we identified several entities that could serve as a basis for implementing some indicative use case scenarios in the virtual space. Using the floor map of the part of the pilot site involved, we started by creating a 3D model of the overall factory site. In the virtual space, the model covers an area of 90 by 40 meters, with 12 meters in height, i.e., it is similar in size with a sector of a typical factory in Europe. We then continued with simple models of the entities chosen to represent important parts of the production line. In this case, we chose 3 cooling tunnels, a number of liquid tanks and a set of mixing machines. We also added typical scenery found in such industrial spaces, in order to have a somewhat more complete ``industrial'' scene available.

This initial 3D model was developed in Blender\footnote{Home of the Blender project, free and open 3D creation software, https://www.blender.org/}, which provides a more sophisticated environment for designing 3D models than typical 3D engines. It also provides advanced capabilities like photorealistic visualization using raytracing, which was helpful in order to create a good-looking prototype in short time. Blender has also the capability to export 3D models in formats understood by Unity, which makes it very easy to import such models and use them directly inside the Unity editor. The wireframe of this initial 3D model inside Blender can be seen in the following figure. 

In Fig.\ref{fig:3d-render-example}, we include a screenshot showing a photorealistic rendering through Blender, showcasing the level of visual fidelity achieved by using rather basic 3D models. This helped relay the design elements to the consortium partners involved in pilot sites during design and implementation in an easy manner, showcase the potential of the overall approach to them and get them involved, as well as kickstart the conversation with them regarding the overall design approach.  

\begin{figure}
\begin{center}
    \includegraphics[width=\columnwidth]{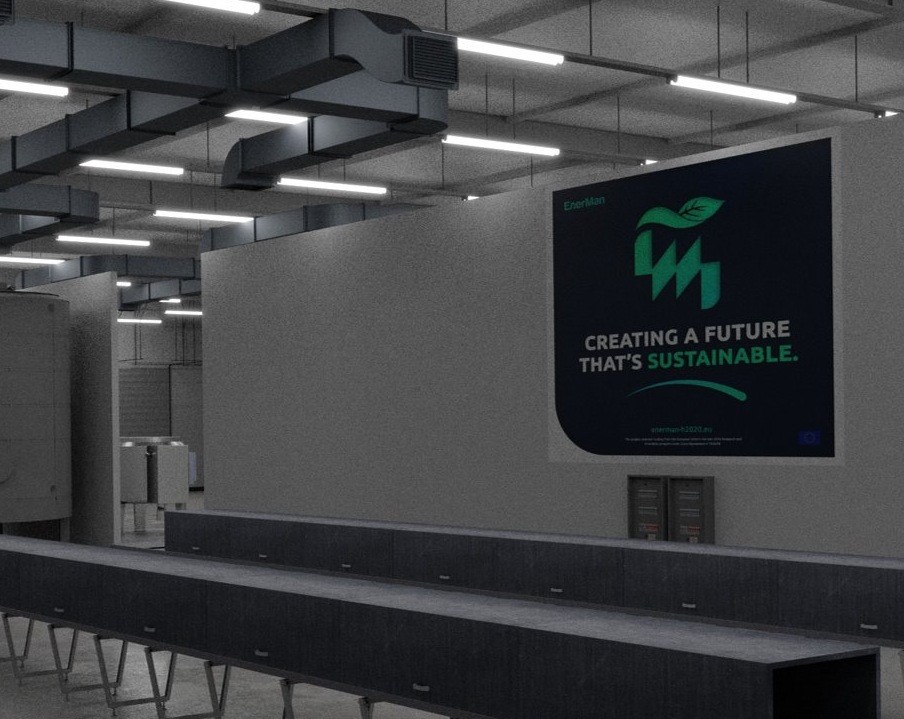}
    \caption{A screenshot from the 3D factory prototype model inside Blender.}
    \label{fig:3d-render-example}
\end{center}
\end{figure}

In this process, we had a series of somewhat structured interviews with ENERMAN pilot sites in order to communicate our designs and get more analytic feedback from them regarding specific aspects that we thought would be directly interesting to them, as well as more general aspects. Although the 3D models were not, at this stage, copies of the actual sites or the machines stationed inside them, they were deemed by these partners as ``good enough'' to relay the message and proceed with the conversation on aspects more directly related to their own sites and use cases. 

After this initial stage of 3D model development, we proceeded with experimenting on defining a ``pipeline'' for building 3D models of pilot sites and specific machinery that would be closer to their real-world counterparts, and not be generic representations of classes of equipment, machinery, etc. For this purpose, we also aimed to use widely available equipment and tools, and not be based on specialized hardware and software platforms, that could make it difficult to scale such processes across several pilot sites due to cost, complexity or incompatibility. Having this in mind, to speed up the process of creating more accurate digital twins in terms of visual representation, we utilized LiDAR-equipped devices to scan areas of the pilot sites and then use these scans to create 3D models that can be directly imported to platforms like Unity. The most promising candidate solution for facilitating this process was to use LiDAR-equipped mobile devices, like smartphones and tablets, that are available in the market at reasonable cost and offer significant capabilities.  In summary, an iPad Pro is used to create point clouds of areas, machines or other assets, which are then processed to clean-up the generated datasets and create simple 3D meshes. These 3D meshes can then be processed in tools like Blender to create more functional 3D models, which can finally be imported to Unity for integration with the actual sites’ digital counterparts.

\section{Conclusions - Future Work}

In recent years, VR and AR have become relatively popular in the context of industrial applications, and we are now beginning to see more focused attempts at utilizing such technologies to make the vision of Industry 4.0 a reality. A continuous stream of better and more affordable hardware platforms for VR and AR experiences in the consumer market has helped to make access to such platforms much easier, and at this stage more sectors of the industry are beginning to show interest in applying AR and VR in their production lines.

Having the above in mind, we presented the design and implementation of a VR/AR toolbox for the ENERMAN project, which provided a basis for implementing a prototype showcasing design elements that will be utilized within future prototypes for the pilots of the project. We outlined its design and implementation, relaying the general philosophy of its core design elements, and discussed several general end-user requirements within ENERMAN. We then delved into the overall architecture produced, describing the main technologies and components involved, as well as presented aspects related to the end-user interfaces. We also discussed aspects of the pipeline used to produce a prototype for showcasing the possibilities enabled by the design presented here.

Moving forward, the design approach and software pipeline presented here will aid in providing a framework driving the development of more elaborate tools for virtual factories and digital twinning within the ENERMAN project. We also aim to implement more elaborate mechanisms for the AR part of the toolbox, such as more fine-grained proximity detection.

\bibliographystyle{IEEEtran}
\bibliography{enerman-bibliography}

\end{document}